\documentclass[sigconf]{acmart}
\usepackage{enumerate}
\usepackage{tabularx}
\usepackage{array,multirow}
\usepackage{float}

\copyrightyear{2022} 
\acmYear{2022} 
\setcopyright{rightsretained} 
\acmConference[GE@ICSE'22]{Third Workshop on Gender Equality, Diversity, and Inclusion in Software Engineering}{May 20, 2022}{Pittsburgh, PA, USA}
\acmBooktitle{Third Workshop on Gender Equality, Diversity, and Inclusion in Software Engineering (GE@ICSE'22), May 20, 2022, Pittsburgh, PA, USA}
\acmDOI{10.1145/3524501.3527595}
\acmISBN{978-1-4503-9294-5/22/05}

\begin{document}

\title{The Role of Diversity in Cybersecurity Risk Analysis: An Experimental Plan}

\author{Katja Tuma}
\affiliation{%
  \institution{Vrije Universiteit Amsterdam}
  \country{}
  }
\email{k.tuma@vu.nl}

\author{Romy Van Der Lee}
\affiliation{%
  \institution{Vrije Universiteit Amsterdam}
  \country{}
  }
\email{r.vander.lee@vu.nl}

\begin{abstract}
Cybersecurity threat and risk analysis (RA) approaches are used to identify and mitigate security risks early-on in the software development life-cycle. 
Existing approaches automate only parts of the analysis procedure, leaving key decisions in \textit{identification, feasibility and risk analysis, and quality assessment} to be determined by expert judgement.
Therefore, in practice teams of experts manually analyze the system design by holding brainstorming workshops.
Such decisions are made in face of uncertainties, leaving room for biased judgement (e.g., preferential treatment of category of experts).
Biased decision making during the analysis may result in \textit{unequal contribution of expertise}, particularly since some diversity dimensions (i.e., gender) are  underrepresented in security teams.
Beyond the work of risk perception of non-technical threats, no existing work has empirically studied the role of diversity in the risk analysis of technical artefacts.
This paper proposes an experimental plan for identifying the key diversity factors in RA.
\end{abstract}




\settopmatter{printfolios=true}
\maketitle

\section{Introduction}
Security-by-design techniques strive to avoid risks before the software is actually developed to reduce the security incurred costs in later stages of development.
For instance, threat and risk analysis (RA, for short) can be performed by adopting systematic techniques to scrutinize the software system (e.g., STRIDE~\cite{shostack2014threat}, attack trees~\cite{saini2008threat}, CORAS~\cite{lund2010model}, to name a few).
Due to development trends (Agile, DevOps), analysis automation is a much needed direction for improvement~\cite{cruzes2018challenges}.
However, many automation efforts automate parts of the analysis procedure and rely on an extended manual modeling effort~\cite{tuma2018threat}.
Key decisions in the \textit{identification, feasibility and risk analysis, and quality assessment} are often left to the experts.
In practice, teams of experts often analyze the system design manually by holding brainstorming workshops where they discuss potential risks and decide which risks will be mitigated and how.
Risk based decisions must be made in face of uncertainties~\cite{bier2020role}, leaving room for biased judgement (e.g., preferential treatment of category of experts).
In addition, the quality of analysis outcomes is ultimately determined by (potentially biased) expert judgement~\cite{jaspersen2015probability,brito2020predicting}.
For instance, evidence of groupthink~\cite{wang2018groupthink} in safety analysis sessions suggests that knowledge is \textit{not always contributed (equally)} by all participants, but exact variables that effect the technical quality of the analysis are not yet understood.
Further, the type of judgement bias present in risk-based decision making and more generally the role of diversity has not yet been sufficiently explored in RA (as discussed in the related work, Section~\ref{sec:relatedwork}).

In many engineering domains, there is a lack of security experts~\cite{tuma2021finding}.
In addition, security expert teams are not diverse, especially for what concerns gender. According to a news post by the International Consortium of Minority Cyber Professionals (non-profit) in August 2021~\cite{Cyversity:web}, the cybersecurity workforce in the USA is currently only 14\% female. 
Generally speaking, there is a significant mismatch between the cultural notion of a successful technical professional and the perceived skills that are needed for success in the computer science domain. Wynn and Correll~\cite{wynn2017gendered} show that men are more likely than women to believe they possess the stereotypical traits and skills of a successful tech employee. This mismatch can explain the low technical confidence of young female professionals that are deciding about their career path in computer science~\cite{beyer2003gender}.
We believe there is still untapped potential in empowering women to pursue their career as security experts. 
This motivates us to gather empirical evidence that could help in
(i) raising more interest and confidence levels of female students in cybersecurity topics, and
(ii) encouraging organizations to build gender balanced security teams.
\looseness=-1

The purpose of the laid out research track is to identify the diversity dimensions (see Section~\ref{sec:background}) that play a role during RA of more sophisticated technical artefacts. To simulate RA in the experimental setup, the ACM Code of Ethics scenarios can be used as RA task when interacting with the communication science population as opposed to traditional economics/social experiments. Second, case studies with detailed information about the software design (e.g., IoT home monitoring system used in~\cite{tuma2018two}) can be used as RA task when interacting with the computer science population. The experimental conditions will differ depending on the measure and target audience (as described in Section~\ref{sec:approach}).
This paper discusses the challenges (see Section~\ref{sec:challenges}) and outlines the experimental plan (see Section~\ref{sec:approach}) for identifying the role of diversity in the technical domain of RA.
Our experimental plan is three-fold. First, we plan to characterize the key differentiating factors in diversity dimensions (e.g., gender) that are present to RA. Second, we plan to evaluate their effects in a non-technical population and technical population. Third, we plan to develop and validate diversity interventions (given to participants in the form of training) in realistic (technical) environments.
We briefly mention the next steps in the laid out plan and present the concluding remarks in Section~\ref{sec:conclusion}.

\section{Diversity Dimensions in RA}
\label{sec:background}
Diversity processes are well understood in the social science disciplines, such as, the role of diversity factors in group dynamics or effects of diversity biases in organizational processes.
In contrast to the technical domain, studies have demonstrated how diversity (gender, nationality) can be beneficial for decision making and progress, once effectively incorporated ~\cite{nielsen2017opinion,van2018perceptions,van2012defying}. 

\textbf{Gender.}
Gender is an individual's own gender identity, which is typically, man and woman, but can also be non-binary.
Rodriguez et al.~\cite{rodriguez2021perceived} found evidence of bias against women in some software engineering communities, and sometimes negative perceptions about women working in teams.
Thus, this is an interesting dimension to further investigate in the context of cybersecurity.

\textbf{Education.}
Education is an individuals' achieved level and topic of specialization (e.g., computer security vs AI) of academic studies. 
Risk-based decisions have to be made in organizations by the managerial layers, who typically have a good understanding of the product, but do not necessarily posses the technical skills of security experts or engineers.
Therefore, it is interesting to investigate this dimension and include participants from a different domain (e.g., with background in communication sciences).
Education turned out to be a non-significant variable in the study of the impact of commercial Antivirus on people’s awareness of security incidents~\cite{jardine2020case}.
However, it is not clear whether this dimension has an impact in performing a RA task.

\textbf{Nationality \& Race.}
Nationality is the country of origin, which is often coupled with the culture and language that categorizes social groups.
Race is a social construct linked with individual's physical characteristics such as skin color and is used to categorize populations.
Determining the effect of racial (and nationality) bias in cybersecurity practices is to date an open question.
Thomas et al.~\cite{thomas2018speaking} conducted semi-structured interviews with 14 Black women in computing and report that Black women experienced isolation (though it is not clear whether due to gender or race).
But, few studies have focused on racial (and nationality) diversity in the software engineering discipline~\cite{rodriguez2021perceived}.

\textbf{Age \& Seniority.}
Age and seniority (i.e., position in the organization and years of experience) may have an impact on both the technical outcomes of a RA task and on the judgment of outcomes quality (e.g., over-favoring results submitted by senior analysts).
For instance, a category of experts (e.g., junior vs senior) may underestimate the ease with which a mitigation can be implemented in reality~\cite{wright2002empirical}. 
This may hurt later-on when infeasible mitigations have to be replaced with cheaper (and perhaps) less insecure solutions.
We remark that though this dimension is indeed relevant, empirically observing seniority (in a realistic setting) requires involving professionals. Our experimental plan is designed in an academic (controlled) setting with student participants. Thus we do not attempt to observe this dimension as part of this work.

\section{Related Work}
\label{sec:relatedwork}
There is an extensive account of existing RA techniques, their main characteristics, information sources, and limitations. We refer the interested reader to the literature reviews by Tuma et al.~\cite{tuma2018threat} and Ling et al.~\cite{ling2020systematic}.
For an overview of research on diversity theory and the relevant social processes involved in organizations (e.g., group dynamics) see studies~\cite{nielsen2017opinion,van2018perceptions,van2012defying}.

\subsection{Diversity Studies in Risk Analysis}
Beyond the gender theory about different risk perception~\cite{gustafsod1998gender}, there was little scientific inquiry in the computer science domain about gender diversity (or other diversity processes) in risk analysis in the past 10 years. 
When considering non-technical scenarios (e.g., returning from work at night or natural disasters) different risk perceptions between men and women (and ethnic minorities) have been investigated, mainly in the USA but also in Sweden~\cite{olofsson2011white}.
While the studies conducted in the USA suggest that white males perceived lower risk compared to women and minorities, the same was not observed by Olofsson and Rashid for Sweden~\cite{olofsson2011white}. 
A recent study~\cite{jardine2020case} investigated the impact of commercial Antivirus on people’s awareness of security incidents with 1000 respondents. They found that the only relevant variables were online insecurity, risk aversion and \textit{gender} (other non-significant variables were education, age, experience with information systems).
Other empirical studies~\cite{fenwick2001effect,myaskovsky2005effects} claim a diversity based on participation in simple games.
\looseness=-1

Thus, we discern that existing research has focused on measuring diversity aspects in RA when participants consider non-technical artefacts while there is still the need to study diversity factors when considering technical (e.g., cybersecurity threats and mitigations) artefacts.
\looseness=-1

\subsection{Diversity Studies in Computer Science \& Software Engineering}
Rodriguez et al.~\cite{rodriguez2021perceived} have recently conducted a systematic literature review to identify the diversity aspects that have been studied in software engineering and development. 
Gender has been the most widely researched dimension (61\% of the papers in~\cite{rodriguez2021perceived}), while others (nationality, age, race) are less explored. 
This does not come as a surprise, considering that computer science, engineering, and physics are less gender balanced compared to other STEM fields (e.g., biology)~\cite{cheryan2017some}.
In addition, existing research has focused on studying diversity in the context of open source communities and developers (e.g., see the study on gender bias on GitHub~\cite{imtiaz2019investigating}), while other domains are less explored.
For instance the role of gender-diversity in the domain of software architecture is unexplored~\cite{razavian2015feminine,spichkova2017role}.
Some tools have been proposed to assess diversity dimensions of software or software development processes (GenderMag, InclusiveMag, and AID Tool, to name a few).
However, most studies have focused on identifying bias through case studies, rather than conducting experiments or proposing new models, tools, or processes~\cite{rodriguez2021perceived}.

GenderMag~\cite{vorvoreanu2019gender} is a method to inspect software (e.g., interface design) to identify gender-inclusiveness issues. InclusiveMag~\cite{mendez2019gendermag} is a generalization of GenderMag that can be used to generate inclusiveness methods for a particular dimension of diversity.
AID Tool~\cite{9402060} is an automation of the GenderMag method to detect ''inclusivity bugs'' which are software features that disproportionately disadvantage particular groups of contributors in OSS. The tool was evaluated on 20 GitHub projects.



\section{Challenges}
\label{sec:challenges}
This section outlines the challenges in establishing the role of diversity in RA.

\subsection{Evidence of Biased Judgement}
The first challenge is capturing whether diversity dimensions have an effect on biased judgment. 
To observe this we plan to measure the \textit{type} (i.e., gender bias, education bias, and nationality and race bias) \textit{and amount of bias} the analyst may be projecting when making (subjective) judgements about \textit{another analyst's} work (report of outcomes).
We are interested to observe the evidence of biased judgement in studies with and without the technical artefact and in different populations (i.e., computer science and communication science).
Measuring such bias is relevant considering that sometimes a preferential (managerial) choice must be made between analysis reports of two different analysts which may belong to different diversity groups (e.g., man vs woman).
Lastly, capturing the interplay between the evidence of biased judgment and \textit{actual} diversity effects on analysis outcomes is an unexplored direction.
We may find there are no actual differences in the outcomes produced, but there is evidence of biased judgments (or vice-versa).



\subsection{Evidence of Diversity Effects on Outcomes}
\begin{table}[]
    \centering
    \noindent\setlength\tabcolsep{4pt}%
    \begin{tabular}{p{0.3\columnwidth} | p{0.7\columnwidth}}
    \toprule
        \textbf{Outcome} &\textbf{Outcome type} \\
    \midrule
        Threat & spoofing, tampering, repudiation, information disclosure, denial of service, elevation of privilege \\
        Assumption & domain, security \\
        Attack surface & physical, close-proximity, remote \\
        Effected security objective & confidentiality, integrity, availability, accountability \\
        Threat impact & financial, safety, operational, reputation, legal (incl. privacy)\\
        Risk priority & high, medium, low \\
        Threat mitigation & preventative, detective/reactive, corrective \\
    \bottomrule 
    \end{tabular}
    \caption{Type of RA outcomes investigated in this work}
    \label{tab:outcome-types}
\end{table}

The second challenge is to investigate whether diversity dimensions have an \textit{actual} effect on the task of RA.
To this aim, we will measure the effects of diversity dimensions on the RA outcomes.
Since the quality of analysis lacks a formalised definition (e.g., often natural language is used to describe attack scenarios and informal notations are used for modeling~\cite{tuma2018threat}), we will use measures that can be easily reproduced.
Namely, we can observe how diversity dimensions effect the \textit{type of analysis outcomes}.
Table \ref{tab:outcome-types} shows various outcomes that are often included in the analysis report. In what follows, we describe the outcomes and motivate their inclusion for observation.
\looseness=-1

\textbf{Threats.} A malicious attacker (with means and motivations) poses a threat to harm the desired security properties of system assets.
As part of the analysis outcomes, the \textit{identified threats} (e.g., spoofing an authorized user) are accompanied by a detailed description of the attack scenarios (e.g., CAPEC-656: Voice Phishing~\footnote{https://capec.mitre.org/data/definitions/656.html}) that realize the threat.
We borrow the threat categories from the documentation of STRIDE, a threat modeling methodology developed by Microsoft~\cite{shostack2014threat}.

Analyses conducted by experts tend to be more balanced in terms of the categories of identified threats, whereas RA novices tend to report disproportionately more tampering, denial of service and information disclosure threats~\cite{tuma2018two,tuma2021finding}.
We are interested to observe whether category distribution patterns emerge for other diversity dimensions.

\textbf{Assumptions.} Assumptions are statements about the system under analysis that may or may not be true~\cite{tuma2021finding}. While system requirements and constraints are typically explicitly defined and static, assumptions are often implicit and dynamic in nature (i.e., they can be invalidated and modified as the project evolves). 
Due to little support offered to the analysts regarding the assumption definition, quality assessment, and management, a recent study investigated the role of assumptions in context of STRIDE~\cite{van2021descriptive}.
Van Landuyt and Joosen~\cite{van2021descriptive} find that the majority of assumptions (created by students during STRIDE) were used to either justify an existence of threats or are used to eliminate threats.
In~\cite{van2021descriptive} a substantial subset (78\%) of the assumptions was in direct reference to security-related concepts (i.e., security assumptions), however also domain assumptions (statements about component functionalities) were made.

In the safety science domain, managerial over-reliance on routine technical inspection was identified as a common underlying characteristic of human-created disasters~\cite{waring2015managerial}. In RA, a category of experts may make assumptions about different properties of the system and similar over-reliance on verifying assumptions could be present. We are interested to investigate the effect of other diversity dimensions on the type of assumptions that are made during RA.

\textbf{Attack surface.} Security analysts often rely on defining the \textit{attack surface} and the required attacker profile to exploit it to determine the feasibility of an attack scenario.
Intuitively, consider the feasibility of an information disclosure threat on internal communication channels to expose intercepted emails by a 'script kiddie' with little computing resources and off-the-shelf hacking tools.
The attack may require gaining physical access to the organizations internal network and breaking TLS encryption.
In addition, the 'script kiddie' may not have a motive to launch this attack.
Determining feasibility is subjective, domain-specific and not always obvious.

Giddens et al.~\cite{giddens2020gender} found that when analyzing internal threats, managers will consider male employees as exhibiting greater malicious intentions (associated with computer abuse) compared to female employees).
Thus, it is interesting to measure whether diversity dimensions play a role in identifying the core variables for determining threat feasibility.

\textbf{Effected security objective.} The attacker goal is to compromise a \textit{security objective} (confidentiality, integrity, availability, and accountability) of an asset of value.
Since spoofing and elevation of privilege threats may compromise multiple security objectives (e.g., a lack of authentication due to spoofing may result in loss of confidentiality, integrity, or availability), we are also interested to observe possible patterns of the threatened security objectives per diversity factor. 

\textbf{Threat impact.} Given a successful attack, threat impact refers to the damage it causes to the organization. 
Estimating the type (and amount) of impact is useful for organizations to put the threat into perspective of the domain (e.g., when safety-critical systems are analyze, the safety impact can not be left unmitigated).
Some impacts may be easier to estimate depending on the background and previous education. 
For instance, for an engineer the number of effected units may seem easier to determine compared to legal and financial loss estimates, for which other department expertise may be needed~\cite{cruzes2018challenges}.
Thus, we are interested to investigate whether diversity dimensions (e.g., education) have an effect on the type of impacts that our participants identify.

\textbf{Risk priority.} Since the number of identified threats quickly explodes in realistic projects~\cite{tuma2021finding}, they must be prioritized based on estimations of risk (e.g., the last step of STRIDE~\cite{shostack2014threat}).
The most critical threats are red flags and consequently larger investments are made to develop and test their mitigations.
We refer to risk as a product of threat probability and impact.
How individuals assess risk priorities may be related to their risk perception which is already well understood~\cite{gustafsod1998gender}.

\textbf{Mitigations.} Mitigations of a cybersecurity risk can be preventative (e.g., implementation of two-factor authentication), detective/reactive (e.g., using intrusion detection and access revocation techniques) and corrective (such as maintaining audit trails or restoring from a secure state).
Multiple strategies can be adopted to counter a cybersecurity threat, and the final choice may depend on domain-related factors, as the cost of implementing the mitigation has to be reasonable.
A category of experts (e.g., man vs woman) may underestimate the ease with which a mitigation is actually implemented, as observed in~\cite{wright2002empirical}. By doing so, they in fact contribute to the actual mitigation of the threat.
Thus, we are interested to observe how diversity dimensions effect the type of mitigations that are identified during the analysis.

\subsection{Identification of countermeasures to identified effects}
Because risk decisions must be made in face of uncertainties for which no statistical bias is available~\cite{bier2020role}, organizational issues play a major role in which the final technical artefact is chosen~\cite{pence2020discourse} (i.e., security mitigations), and thus we must put in place technical/organizational measures to counterbalance it. 
Our final challenge is to identify:
\begin{enumerate}[i]
    \item RA practices (e.g., methods, tools, processes borrowed from adjacent disciplines) that help foster diversity, and
    \item feasible and effective countermeasures for organizations to adopt.
\end{enumerate}
Identifying diversity fostering practices could be partially achieved by conducting a more systematic review of the existing literature.
However, to evaluate the identified countermeasures empirically more future efforts will be needed. 

\section{Experimental Plan}
\label{sec:approach}

\textbf{Study of diversity effects and their prevalence in RA.} 
There is a fairly complete account and understanding of the existing RA techniques, their main characteristics, information sources, and limitations~\cite{tuma2018threat,ling2020systematic}. 
On the other hand, research has shown, with regard to gender diversity, that there are differences in the perceived competence of men and women~\cite{van2018perceptions}. In the context of academic funding competitions, to illustrate this process, there is evidence that men and women produce ideas of similar quality, yet women are underestimated for their competence~\cite{van2015gender}. In other words, this implies that women are likely to be equally competent in identifying security threats as men, yet their assessments might not be sufficiently valued. Hence, this might have far reaching consequences for the RA as this might result in an underestimation of the security threats. It is thus pivotal to investigate whether diversity processes are at play during RA. 

\smallbreak
\noindent
\textbf{RQ1.} \textit{What diversity processes play a role during security threat analysis and risk assessment of IT systems?} 
\smallbreak

\textbf{Experimentation and validation of diversity interventions in RA.} Previous studies have experimented with RA techniques with practitioner and student participants ~\cite{tuma2018two,tuma2021finding,labunets2013experimental,labunets2017equivalence} measuring (among others) technique productivity, analysis outcomes, perceived ease of use.
But, diversity relevant measures, controls and relevant training for RA techniques have not yet been defined for such experiments~\cite{rodriguez2021perceived}. 
In contrast, Van der Lee and Ellemers~\cite{van2018perceptions} have identified characteristics of effective diversity interventions, and applied those to practical decision making processes. 
One of the key features of diversity intervention effectiveness is the degree to which they are applicable to the current situation. In other words, whether participants can use the knowledge that they gained~\cite{homan2015interplay}. The interventions thus need to promote ways in which analysts can use their diversity to better their decision making. A factor that might hinder this process is the perceived objectivity of the analysts, thereby paradoxically increasing rather than decreasing implicit biases rooted in stereotypes~\cite{uhlmann2007think}. Hence, to strengthen the quality of RA the diversity interventions must be grounded in theory and tested in the context of RA. 

\smallbreak
\noindent
\textbf{RQ2.} \textit{What experimental protocols, metrics, and controls capture and effectively incorporate diversity in the context of RA of IT systems?}
\smallbreak

\subsection{Hypotheses \& Methods}

\begin{figure}
    \centering
    \includegraphics[width=\columnwidth]{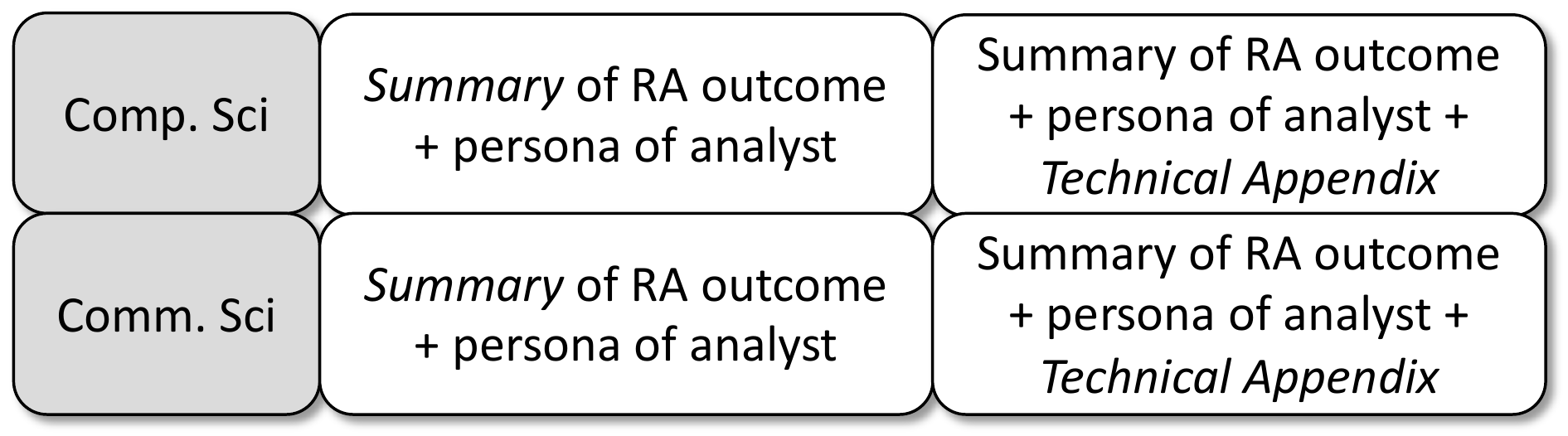}
    \caption{Design of the experiment with two versions of the survey for computer science (Comp. Sci) and communication science (Comm. Sci) populations}
    \label{fig:surveys}
\end{figure}

To answer the research questions RQ1 and RQ2, we formulate two type of hypotheses. 
First, we pose three hypotheses about the \textbf{equivalence of the sample means for biased judgement} (e.g., by using Two One-Sided T-Tests (TOST)~\cite{meyners2012equivalence}) in presence and absence of the technical artefact.

To test hypotheses $H_1-H_3$, we designed a survey experiment with two populations, computer science BSc students and communication science BSc students.
The survey is targeting students taking courses taught by the experimenters.
Individual participation to the survey will be voluntary.
The survey will be based on a high-level scenario (such as the ACM ethics Malware Disruption case~\footnote{https://ethics.acm.org/code-of-ethics/using-the-code/case-malware-disruption/}) and will require the participants to understand basic principles of security and privacy.
We adopted a blocked design for the experiment with two versions of the survey which are distributed to both populations, as shown in Figure~\ref{fig:surveys}.
The first will include a high-level \textit{executive summary} of the RA outcomes (in security layman terms) including the final recommendation (i.e., threat mitigation) and a \textit{fictitious persona} description of the RA analyst.
The fictitious personas (also called vignettes) can be used to measure biased judgements.
The second survey version will include the executive summary, the final RA recommendation, the fictitious persona, and a \textit{technical appendix} with the actual RA outcomes in full detail.
Table~\ref{tab:vignettes} shows the dependent variables used to measure the biased judgement. The level of each vignette dimension (gender, seniority, and mitigation) is randomly sampled and presented to the participant (similar to survey experiment in~\cite{hibshi2015assessment}) in a context of the ACM Code of Ethics scenario. The participant will be then asked to rate their trust in the analyst and agreement with the suggestion of the analyst. Several control questions will also follow. The independent variables are the participant gender and education background.

\begin{table}[]
    \centering
    \footnotesize
    \noindent\setlength\tabcolsep{4pt}%
    \begin{tabular}{p{0.2\columnwidth}  p{0.75\columnwidth}}
    \toprule
        Dimension & Level \\
    \midrule
        $\$Name$ & $\bullet$ (male) Frank \\
         & $\bullet$ (female) Ana \\
    \midrule
        $\$Seniority$ & $\bullet$ (senior) Senior Analyst\\
         & $\bullet$ (junior) Junior Analyst\\
    \midrule
        $\$Mitigation$ & $\bullet$ (detective/reactive) coordinating the web browsers with blacklists which block any incoming traffic from server hosting malicious activity. \\
         & $\bullet$ (corrective) engineering a worm which targets behavior resembling “hacker” activity on the effected server.  \\
    \bottomrule 
    \end{tabular}
    \caption{The vignette dimensions and levels for the survey experiment designed to measure bias in the judgement of RA outcomes.}
    \label{tab:vignettes}
\end{table}

\smallbreak
\noindent
$ H_1: Comp. Sci = Comm. Sci$
\smallbreak

\noindent
First, we will measure whether the judgement bias (i.e., the type of bias and the effect size) in the computer science population is equivalent to the judgement bias in the communication science population in absence of the technical artefact.

\smallbreak
\noindent
$ H_2: Comm. Sci + Artefact = Comm. Sci$
\smallbreak

\noindent
If we find that the bias is in fact equivalent across the two domains, it is interesting to also investigate the equivalence of judgement bias for the communication science population with the technical artefact compared to the communication science population without the technical artefact.

\smallbreak
\noindent
$ H_3: Comp. Sci + Artefact <> Comp. Sci$
\smallbreak

\noindent
If we find that the bias is in fact equivalent within the non-technical domain, we must (naturally) also find bias equivalence within the technical domain.
Therefore, it is interesting to investigate the equivalence of judgement bias for the computer science population with the technical artefact compared to the computer science population without the technical artefact. 
If our statistical analysis does not support the equivalence hypothesis, 
we must further investigate how the two samples differ.
For instance, if compared to the computer science population without the technical artefact the judgement bias of the computer science population with the technical artefact is greater, this is evidence of a systematic bias in the computer science community.



\begin{figure}
  \includegraphics[width=\columnwidth]{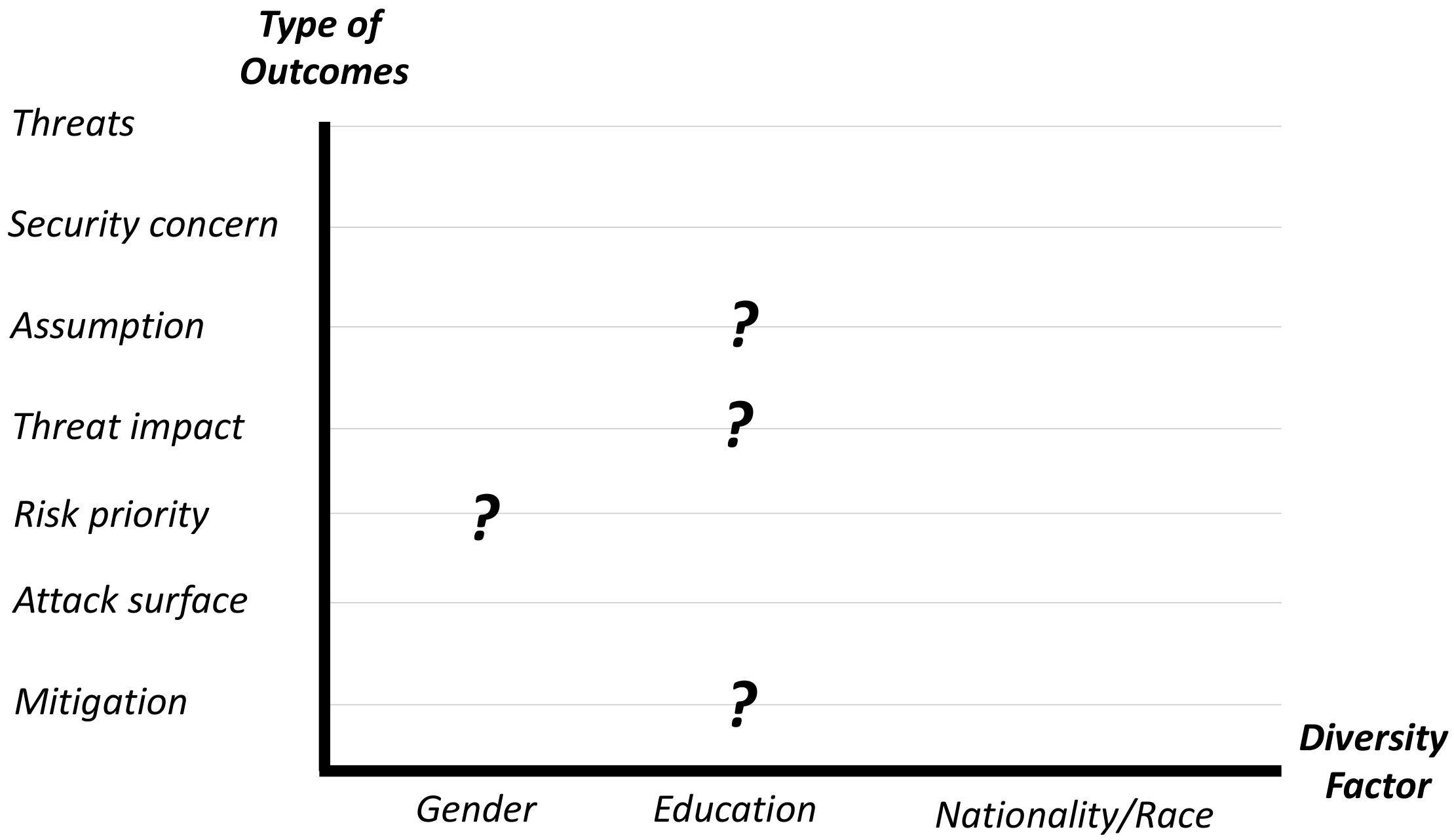}
  \caption{Possible (?) diversity effects and expected absence of effects in RA outcome type studies in presence of the technical artefact}
  \label{fig:hypotheses-cs}
\end{figure}

Second, we pose hypotheses about the \textbf{equivalence of the sample means for the RA outcomes} in presence of the technical artefact.
Figure~\ref{fig:hypotheses-cs} depicts hypotheses $H_4:H_6$.
To test hypotheses $H_4-H6$ we will conduct controlled experiments with computer science MSc students.
Concretely, we will hand out the documentation of a software system (e.g., requirements, component and deployment diagrams etc.) to individual students. 
After a training session, the students will be tasked to perform a RA following a prescribed technique and hand-in the RA outcomes.

\bigbreak
\noindent
$H_4: Comp. Sci F = Comp. Sci M$
\smallbreak

\noindent
Regarding gender, we expect that the outcome types reported by women are equivalent to the outcome types reported by men. 
Studies of risk perception suggest that women perceive certain risks differently compared to men.
Though we do not foresee strong differences, we might find some effects when it comes to risk priority.
\looseness=-1

\smallbreak
\noindent
$H_5: Comp. Sci 1 = Comp. Sci 2$
\smallbreak

\noindent
Regarding education, we expect that the students of various specialization tracks report equivalent outcome types for the same system under analysis.
Our population is computer science students, with some differences in the elective courses and program choices (e.g., we plan to include students from various master programs, such as IA, computer Security, and Software Engineering).
We may find some outliers in the proportions of domain vs security assumptions made, depending on the previous completed courses or the master specialization. 
For example, possibly AI students make less security assumptions, compared to Computer Security students.
Though we do not have string convictions, we may find that previous courses may have an effect on the threat impact (e.g., students with robotics and AI background may focus on safety impacts).
Similarly, we may find come differences in the type of mitigations considered by students of different background.


\smallbreak
\noindent
$H_6: Comp. Sci Ra = Comp. Sci Rb$
\smallbreak

\noindent
We expect that the students of various race and nationality report statistically equivalent outcome types. We have not found any research that would suggest otherwise.
\looseness=-1

The findings from investigating $H1 - H6$ will play a key role in developing diversity interventions in the form of training. To validate the diversity training, we plan to conduct comparative experiments in a similar controlled setting, where we administer an additional diversity training to half of the participants. 

\section{Potential Threats to Validity}
\label{sec:threats}
\textbf{Small female population.}
There is typically around 20\% (or less) female students enrolled in computer science programs.
We are aware of the validity threats caused by an unbalanced population sample, which is omnipresent in all gender diversity studies in STEM disciplines~\cite{rodriguez2021perceived}.
To partially mitigate this threat, we planned a sister experiment (with a less technically demanding task) with a communication science population (see Section~\ref{sec:approach}).
Second, we will rally female computer scientist students towards participation through local steminist groups and similar community organized channels.

The small female population in computer science is also the rationale for organizing the controlled experiments with computer science population (testing hypotheses $H_4-H6$) with individual students.
A more realistic setup (as was done in previous studies~\cite{tuma2018two}) would include dividing participants into teams as in practice RA is performed in teams of experts.
But measuring the individual female contribution to the outcomes inside gender-mixed teams becomes very hard and possibly not reliable.


\textbf{Generalizability of results to practitioners.} We do not plan to include practitioners experienced in RA to participate in our studies.
This limits us in observing the full complexity of the diversity effects (e.g., including seniority) that are actually present in organizations where RA is routinely performed.
Still, studies have shown~\cite{runeson2003using, host2000using, salman2015students} that the differences between the performance of professionals and graduate students are often limited.
To extend our observations we plan to conduct industrial cases studies in the future (e.g., similar to the study performed by Wang and Wagner~\cite{wang2018groupthink}).

\section{Conclusion}
\label{sec:conclusion}
This paper outlines an experimental plan to discover the role of diversity in cybersecurity risk analysis. We present the relevant diversity dimensions that will be observed and discuss the challenges of capturing evidence about biased judgement during RA and about diversity effects on the actual RA outcomes. As a first step, we propose to measure the key differentiating diversity factors in RA by conducting survey experiments with a technical and non-technical population. Second, we plan to conduct controlled experiments with the computer science population to record diversity effects on the outcomes and to validate diversity interventions. Finally, an interview study with practitioners would be beneficial for identifying the most urgent diversity dimensions to be addressed first with the interventions.

\begin{acks}
We express our deepest appreciation to prof. Fabio Massacci for his insightful suggestions for the experimental plan.

This research is partially supported by the Network Institute Academy Assistant program (NIAA) at the Vrije Universiteit and by the H2020 AssureMOSS project that received funding from the European Union’s Horizon 2020 research and innovation program under grant agreement No 952647. This paper reflects only the author’s view and the Commission is not responsible for any use that may be made of the information contained therein.
\end{acks}

\bibliographystyle{ACM-Reference-Format}
\bibliography{bibfile}


\begin{thebibliography}{47}


\ifx \showCODEN    \undefined \def \showCODEN     #1{\unskip}     \fi
\ifx \showDOI      \undefined \def \showDOI       #1{#1}\fi
\ifx \showISBNx    \undefined \def \showISBNx     #1{\unskip}     \fi
\ifx \showISBNxiii \undefined \def \showISBNxiii  #1{\unskip}     \fi
\ifx \showISSN     \undefined \def \showISSN      #1{\unskip}     \fi
\ifx \showLCCN     \undefined \def \showLCCN      #1{\unskip}     \fi
\ifx \shownote     \undefined \def \shownote      #1{#1}          \fi
\ifx \showarticletitle \undefined \def \showarticletitle #1{#1}   \fi
\ifx \showURL      \undefined \def \showURL       {\relax}        \fi
\providecommand\bibfield[2]{#2}
\providecommand\bibinfo[2]{#2}
\providecommand\natexlab[1]{#1}
\providecommand\showeprint[2][]{arXiv:#2}

\bibitem[\protect\citeauthoryear{Beyer, Rynes, Perrault, Hay, and Haller}{Beyer
  et~al\mbox{.}}{2003}]%
        {beyer2003gender}
\bibfield{author}{\bibinfo{person}{Sylvia Beyer}, \bibinfo{person}{Kristina
  Rynes}, \bibinfo{person}{Julie Perrault}, \bibinfo{person}{Kelly Hay}, {and}
  \bibinfo{person}{Susan Haller}.} \bibinfo{year}{2003}\natexlab{}.
\newblock \showarticletitle{Gender differences in computer science students}.
  In \bibinfo{booktitle}{\emph{Proceedings of the 34th SIGCSE technical
  symposium on Computer science education}}. \bibinfo{pages}{49--53}.
\newblock


\bibitem[\protect\citeauthoryear{Bier}{Bier}{2020}]%
        {bier2020role}
\bibfield{author}{\bibinfo{person}{Vicki Bier}.}
  \bibinfo{year}{2020}\natexlab{}.
\newblock \showarticletitle{The Role of Decision Analysis in Risk Analysis: A
  Retrospective}.
\newblock \bibinfo{journal}{\emph{Risk Analysis}} \bibinfo{volume}{40},
  \bibinfo{number}{S1} (\bibinfo{year}{2020}), \bibinfo{pages}{2207--2217}.
\newblock


\bibitem[\protect\citeauthoryear{Brito and Dawson}{Brito and Dawson}{2020}]%
        {brito2020predicting}
\bibfield{author}{\bibinfo{person}{Mario~P Brito} {and} \bibinfo{person}{Ian~GJ
  Dawson}.} \bibinfo{year}{2020}\natexlab{}.
\newblock \showarticletitle{Predicting the validity of expert judgments in
  assessing the impact of risk mitigation through failure prevention and
  correction}.
\newblock \bibinfo{journal}{\emph{Risk analysis}} \bibinfo{volume}{40},
  \bibinfo{number}{10} (\bibinfo{year}{2020}), \bibinfo{pages}{1928--1943}.
\newblock


\bibitem[\protect\citeauthoryear{Chatterjee, Guizani, Stevens, Emard, May,
  Burnett, and Ahmed}{Chatterjee et~al\mbox{.}}{2021}]%
        {9402060}
\bibfield{author}{\bibinfo{person}{Amreeta Chatterjee}, \bibinfo{person}{Mariam
  Guizani}, \bibinfo{person}{Catherine Stevens}, \bibinfo{person}{Jillian
  Emard}, \bibinfo{person}{Mary~Evelyn May}, \bibinfo{person}{Margaret
  Burnett}, {and} \bibinfo{person}{Iftekhar Ahmed}.}
  \bibinfo{year}{2021}\natexlab{}.
\newblock \showarticletitle{AID: An Automated Detector for Gender-Inclusivity
  Bugs in OSS Project Pages}. In \bibinfo{booktitle}{\emph{2021 IEEE/ACM 43rd
  International Conference on Software Engineering (ICSE)}}.
  \bibinfo{pages}{1423--1435}.
\newblock
\urldef\tempurl%
\url{https://doi.org/10.1109/ICSE43902.2021.00128}
\showDOI{\tempurl}


\bibitem[\protect\citeauthoryear{Cheryan, Ziegler, Montoya, and Jiang}{Cheryan
  et~al\mbox{.}}{2017}]%
        {cheryan2017some}
\bibfield{author}{\bibinfo{person}{Sapna Cheryan}, \bibinfo{person}{Sianna~A
  Ziegler}, \bibinfo{person}{Amanda~K Montoya}, {and} \bibinfo{person}{Lily
  Jiang}.} \bibinfo{year}{2017}\natexlab{}.
\newblock \showarticletitle{Why are some STEM fields more gender balanced than
  others?}
\newblock \bibinfo{journal}{\emph{Psychological bulletin}}
  \bibinfo{volume}{143}, \bibinfo{number}{1} (\bibinfo{year}{2017}),
  \bibinfo{pages}{1}.
\newblock


\bibitem[\protect\citeauthoryear{Cruzes, Jaatun, Bernsmed, and
  T{\o}ndel}{Cruzes et~al\mbox{.}}{2018}]%
        {cruzes2018challenges}
\bibfield{author}{\bibinfo{person}{Daniela~Soares Cruzes},
  \bibinfo{person}{Martin~Gilje Jaatun}, \bibinfo{person}{Karin Bernsmed},
  {and} \bibinfo{person}{Inger~Anne T{\o}ndel}.}
  \bibinfo{year}{2018}\natexlab{}.
\newblock \showarticletitle{Challenges and experiences with applying Microsoft
  threat modeling in agile development projects}. In
  \bibinfo{booktitle}{\emph{Proceedings of the Australasian Software
  Engineering Conference (ASWEC)}}. \bibinfo{publisher}{IEEE},
  \bibinfo{pages}{111--120}.
\newblock


\bibitem[\protect\citeauthoryear{Cyversity}{Cyversity}{2021}]%
        {Cyversity:web}
\bibfield{author}{\bibinfo{person}{Cyversity}.}
  \bibinfo{year}{2021}\natexlab{}.
\newblock \bibinfo{title}{The International Consortium of Minority Cyber
  Professionals (non-profit)}.
\newblock \bibinfo{howpublished}{\url{https://www.cyversity.org/}}.
\newblock
\newblock
\shownote{(Accessed on 07/01/2022)}.


\bibitem[\protect\citeauthoryear{Fenwick and Neal}{Fenwick and Neal}{2001}]%
        {fenwick2001effect}
\bibfield{author}{\bibinfo{person}{Graham~D Fenwick} {and}
  \bibinfo{person}{Derrick~J Neal}.} \bibinfo{year}{2001}\natexlab{}.
\newblock \showarticletitle{Effect of gender composition on group performance}.
\newblock \bibinfo{journal}{\emph{Gender, Work \& Organization}}
  \bibinfo{volume}{8}, \bibinfo{number}{2} (\bibinfo{year}{2001}),
  \bibinfo{pages}{205--225}.
\newblock


\bibitem[\protect\citeauthoryear{Giddens, Amo, and Cichocki}{Giddens
  et~al\mbox{.}}{2020}]%
        {giddens2020gender}
\bibfield{author}{\bibinfo{person}{Laurie Giddens}, \bibinfo{person}{Laura~C
  Amo}, {and} \bibinfo{person}{Dianna Cichocki}.}
  \bibinfo{year}{2020}\natexlab{}.
\newblock \showarticletitle{Gender bias and the impact on managerial evaluation
  of insider security threats}.
\newblock \bibinfo{journal}{\emph{Computers \& Security}}  \bibinfo{volume}{99}
  (\bibinfo{year}{2020}), \bibinfo{pages}{102066}.
\newblock


\bibitem[\protect\citeauthoryear{Gustafsod}{Gustafsod}{1998}]%
        {gustafsod1998gender}
\bibfield{author}{\bibinfo{person}{Per~E Gustafsod}.}
  \bibinfo{year}{1998}\natexlab{}.
\newblock \showarticletitle{Gender Differences in risk perception: Theoretical
  and methodological erspectives}.
\newblock \bibinfo{journal}{\emph{Risk analysis}} \bibinfo{volume}{18},
  \bibinfo{number}{6} (\bibinfo{year}{1998}), \bibinfo{pages}{805--811}.
\newblock


\bibitem[\protect\citeauthoryear{Hibshi, Breaux, and Broomell}{Hibshi
  et~al\mbox{.}}{2015}]%
        {hibshi2015assessment}
\bibfield{author}{\bibinfo{person}{Hanan Hibshi}, \bibinfo{person}{Travis~D
  Breaux}, {and} \bibinfo{person}{Stephen~B Broomell}.}
  \bibinfo{year}{2015}\natexlab{}.
\newblock \showarticletitle{Assessment of risk perception in security
  requirements composition}. In \bibinfo{booktitle}{\emph{2015 IEEE 23rd
  International Requirements Engineering Conference (RE)}}. IEEE,
  \bibinfo{pages}{146--155}.
\newblock


\bibitem[\protect\citeauthoryear{Homan, Buengeler, Eckhoff, van Ginkel, and
  Voelpel}{Homan et~al\mbox{.}}{2015}]%
        {homan2015interplay}
\bibfield{author}{\bibinfo{person}{Astrid~C Homan}, \bibinfo{person}{Claudia
  Buengeler}, \bibinfo{person}{Robert~A Eckhoff}, \bibinfo{person}{Wendy~P van
  Ginkel}, {and} \bibinfo{person}{Sven~C Voelpel}.}
  \bibinfo{year}{2015}\natexlab{}.
\newblock \showarticletitle{The interplay of diversity training and diversity
  beliefs on team creativity in nationality diverse teams.}
\newblock \bibinfo{journal}{\emph{Journal of Applied Psychology}}
  \bibinfo{volume}{100}, \bibinfo{number}{5} (\bibinfo{year}{2015}),
  \bibinfo{pages}{1456}.
\newblock


\bibitem[\protect\citeauthoryear{H{\"o}st, Regnell, and Wohlin}{H{\"o}st
  et~al\mbox{.}}{2000}]%
        {host2000using}
\bibfield{author}{\bibinfo{person}{Martin H{\"o}st}, \bibinfo{person}{Bj{\"o}rn
  Regnell}, {and} \bibinfo{person}{Claes Wohlin}.}
  \bibinfo{year}{2000}\natexlab{}.
\newblock \showarticletitle{Using students as subjects—a comparative study of
  students and professionals in lead-time impact assessment}.
\newblock \bibinfo{journal}{\emph{Empirical Software Engineering}}
  \bibinfo{volume}{5}, \bibinfo{number}{3} (\bibinfo{year}{2000}),
  \bibinfo{pages}{201--214}.
\newblock


\bibitem[\protect\citeauthoryear{Imtiaz, Middleton, Chakraborty, Robson, Bai,
  and Murphy-Hill}{Imtiaz et~al\mbox{.}}{2019}]%
        {imtiaz2019investigating}
\bibfield{author}{\bibinfo{person}{Nasif Imtiaz}, \bibinfo{person}{Justin
  Middleton}, \bibinfo{person}{Joymallya Chakraborty}, \bibinfo{person}{Neill
  Robson}, \bibinfo{person}{Gina Bai}, {and} \bibinfo{person}{Emerson
  Murphy-Hill}.} \bibinfo{year}{2019}\natexlab{}.
\newblock \showarticletitle{Investigating the effects of gender bias on
  GitHub}. In \bibinfo{booktitle}{\emph{2019 IEEE/ACM 41st International
  Conference on Software Engineering (ICSE)}}. IEEE, \bibinfo{pages}{700--711}.
\newblock


\bibitem[\protect\citeauthoryear{Jardine}{Jardine}{2020}]%
        {jardine2020case}
\bibfield{author}{\bibinfo{person}{Eric Jardine}.}
  \bibinfo{year}{2020}\natexlab{}.
\newblock \showarticletitle{The Case against Commercial Antivirus Software:
  Risk Homeostasis and Information Problems in Cybersecurity}.
\newblock \bibinfo{journal}{\emph{Risk Analysis}} \bibinfo{volume}{40},
  \bibinfo{number}{8} (\bibinfo{year}{2020}), \bibinfo{pages}{1571--1588}.
\newblock


\bibitem[\protect\citeauthoryear{Jaspersen and Montibeller}{Jaspersen and
  Montibeller}{2015}]%
        {jaspersen2015probability}
\bibfield{author}{\bibinfo{person}{Johannes~G Jaspersen} {and}
  \bibinfo{person}{Gilberto Montibeller}.} \bibinfo{year}{2015}\natexlab{}.
\newblock \showarticletitle{Probability elicitation under severe time pressure:
  A rank-based method}.
\newblock \bibinfo{journal}{\emph{Risk Analysis}} \bibinfo{volume}{35},
  \bibinfo{number}{7} (\bibinfo{year}{2015}), \bibinfo{pages}{1317--1335}.
\newblock


\bibitem[\protect\citeauthoryear{Labunets, Massacci, and Paci}{Labunets
  et~al\mbox{.}}{2017}]%
        {labunets2017equivalence}
\bibfield{author}{\bibinfo{person}{Katsiaryna Labunets}, \bibinfo{person}{Fabio
  Massacci}, {and} \bibinfo{person}{Federica Paci}.}
  \bibinfo{year}{2017}\natexlab{}.
\newblock \showarticletitle{On the equivalence between graphical and tabular
  representations for security risk assessment}. In
  \bibinfo{booktitle}{\emph{International Working Conference on Requirements
  Engineering: Foundation for Software Quality}}. Springer,
  \bibinfo{pages}{191--208}.
\newblock


\bibitem[\protect\citeauthoryear{Labunets, Massacci, Paci,
  et~al\mbox{.}}{Labunets et~al\mbox{.}}{2013}]%
        {labunets2013experimental}
\bibfield{author}{\bibinfo{person}{Katsiaryna Labunets}, \bibinfo{person}{Fabio
  Massacci}, \bibinfo{person}{Federica Paci}, {et~al\mbox{.}}}
  \bibinfo{year}{2013}\natexlab{}.
\newblock \showarticletitle{An experimental comparison of two risk-based
  security methods}. In \bibinfo{booktitle}{\emph{Proceedings of the
  International Symposium on Empirical Software Engineering and Measurement}}.
  IEEE, \bibinfo{pages}{163--172}.
\newblock


\bibitem[\protect\citeauthoryear{Ling, Lagerstr{\"o}m, and Ekstedt}{Ling
  et~al\mbox{.}}{2020}]%
        {ling2020systematic}
\bibfield{author}{\bibinfo{person}{Engla Ling}, \bibinfo{person}{Robert
  Lagerstr{\"o}m}, {and} \bibinfo{person}{Mathias Ekstedt}.}
  \bibinfo{year}{2020}\natexlab{}.
\newblock \showarticletitle{A systematic literature review of information
  sources for threat modeling in the power systems domain}. In
  \bibinfo{booktitle}{\emph{International Conference on Critical Information
  Infrastructures Security}}. Springer, \bibinfo{pages}{47--58}.
\newblock


\bibitem[\protect\citeauthoryear{Lund, Solhaug, and St{\o}len}{Lund
  et~al\mbox{.}}{2010}]%
        {lund2010model}
\bibfield{author}{\bibinfo{person}{Mass~Soldal Lund},
  \bibinfo{person}{Bj{\o}rnar Solhaug}, {and} \bibinfo{person}{Ketil
  St{\o}len}.} \bibinfo{year}{2010}\natexlab{}.
\newblock \bibinfo{booktitle}{\emph{Model-driven risk analysis: the CORAS
  approach}}.
\newblock \bibinfo{publisher}{Springer Science \& Business Media}.
\newblock


\bibitem[\protect\citeauthoryear{Mendez, Letaw, Burnett, Stumpf, Sarma, and
  Hilderbrand}{Mendez et~al\mbox{.}}{2019}]%
        {mendez2019gendermag}
\bibfield{author}{\bibinfo{person}{Christopher Mendez}, \bibinfo{person}{Lara
  Letaw}, \bibinfo{person}{Margaret Burnett}, \bibinfo{person}{Simone Stumpf},
  \bibinfo{person}{Anita Sarma}, {and} \bibinfo{person}{Claudia Hilderbrand}.}
  \bibinfo{year}{2019}\natexlab{}.
\newblock \showarticletitle{From GenderMag to InclusiveMag: An inclusive design
  meta-method}. In \bibinfo{booktitle}{\emph{2019 IEEE Symposium on Visual
  Languages and Human-Centric Computing (VL/HCC)}}. IEEE,
  \bibinfo{pages}{97--106}.
\newblock


\bibitem[\protect\citeauthoryear{Meyners}{Meyners}{2012}]%
        {meyners2012equivalence}
\bibfield{author}{\bibinfo{person}{Michael Meyners}.}
  \bibinfo{year}{2012}\natexlab{}.
\newblock \showarticletitle{Equivalence tests--A review}.
\newblock \bibinfo{journal}{\emph{Food quality and preference}}
  \bibinfo{volume}{26}, \bibinfo{number}{2} (\bibinfo{year}{2012}),
  \bibinfo{pages}{231--245}.
\newblock


\bibitem[\protect\citeauthoryear{Myaskovsky, Unikel, and Dew}{Myaskovsky
  et~al\mbox{.}}{2005}]%
        {myaskovsky2005effects}
\bibfield{author}{\bibinfo{person}{Larissa Myaskovsky}, \bibinfo{person}{Emily
  Unikel}, {and} \bibinfo{person}{Mary~Amanda Dew}.}
  \bibinfo{year}{2005}\natexlab{}.
\newblock \showarticletitle{Effects of gender diversity on performance and
  interpersonal behavior in small work groups}.
\newblock \bibinfo{journal}{\emph{Sex Roles}} \bibinfo{volume}{52},
  \bibinfo{number}{9} (\bibinfo{year}{2005}), \bibinfo{pages}{645--657}.
\newblock


\bibitem[\protect\citeauthoryear{Nielsen, Alegria, B{\"o}rjeson, Etzkowitz,
  Falk-Krzesinski, Joshi, Leahey, Smith-Doerr, Woolley, and
  Schiebinger}{Nielsen et~al\mbox{.}}{2017}]%
        {nielsen2017opinion}
\bibfield{author}{\bibinfo{person}{Mathias~Wullum Nielsen},
  \bibinfo{person}{Sharla Alegria}, \bibinfo{person}{Love B{\"o}rjeson},
  \bibinfo{person}{Henry Etzkowitz}, \bibinfo{person}{Holly~J Falk-Krzesinski},
  \bibinfo{person}{Aparna Joshi}, \bibinfo{person}{Erin Leahey},
  \bibinfo{person}{Laurel Smith-Doerr}, \bibinfo{person}{Anita~Williams
  Woolley}, {and} \bibinfo{person}{Londa Schiebinger}.}
  \bibinfo{year}{2017}\natexlab{}.
\newblock \showarticletitle{Opinion: Gender diversity leads to better science}.
\newblock \bibinfo{journal}{\emph{Proceedings of the National Academy of
  Sciences}} \bibinfo{volume}{114}, \bibinfo{number}{8} (\bibinfo{year}{2017}),
  \bibinfo{pages}{1740--1742}.
\newblock


\bibitem[\protect\citeauthoryear{Olofsson and Rashid}{Olofsson and
  Rashid}{2011}]%
        {olofsson2011white}
\bibfield{author}{\bibinfo{person}{Anna Olofsson} {and} \bibinfo{person}{Saman
  Rashid}.} \bibinfo{year}{2011}\natexlab{}.
\newblock \showarticletitle{The white (male) effect and risk perception: can
  equality make a difference?}
\newblock \bibinfo{journal}{\emph{Risk Analysis: An International Journal}}
  \bibinfo{volume}{31}, \bibinfo{number}{6} (\bibinfo{year}{2011}),
  \bibinfo{pages}{1016--1032}.
\newblock


\bibitem[\protect\citeauthoryear{Pence and Mohaghegh}{Pence and
  Mohaghegh}{2020}]%
        {pence2020discourse}
\bibfield{author}{\bibinfo{person}{Justin Pence} {and} \bibinfo{person}{Zahra
  Mohaghegh}.} \bibinfo{year}{2020}\natexlab{}.
\newblock \showarticletitle{A Discourse on the Incorporation of Organizational
  Factors into Probabilistic Risk Assessment: Key Questions and Categorical
  Review}.
\newblock \bibinfo{journal}{\emph{Risk Analysis}} \bibinfo{volume}{40},
  \bibinfo{number}{6} (\bibinfo{year}{2020}), \bibinfo{pages}{1183--1211}.
\newblock


\bibitem[\protect\citeauthoryear{Razavian and Lago}{Razavian and Lago}{2015}]%
        {razavian2015feminine}
\bibfield{author}{\bibinfo{person}{Maryam Razavian} {and}
  \bibinfo{person}{Patricia Lago}.} \bibinfo{year}{2015}\natexlab{}.
\newblock \showarticletitle{Feminine expertise in architecting teams}.
\newblock \bibinfo{journal}{\emph{IEEE Software}} \bibinfo{volume}{33},
  \bibinfo{number}{4} (\bibinfo{year}{2015}), \bibinfo{pages}{64--71}.
\newblock


\bibitem[\protect\citeauthoryear{Rodr{\'\i}guez-P{\'e}rez, Nadri, and
  Nagappan}{Rodr{\'\i}guez-P{\'e}rez et~al\mbox{.}}{2021}]%
        {rodriguez2021perceived}
\bibfield{author}{\bibinfo{person}{Gema Rodr{\'\i}guez-P{\'e}rez},
  \bibinfo{person}{Reza Nadri}, {and} \bibinfo{person}{Meiyappan Nagappan}.}
  \bibinfo{year}{2021}\natexlab{}.
\newblock \showarticletitle{Perceived diversity in software engineering: a
  systematic literature review}.
\newblock \bibinfo{journal}{\emph{Empirical Software Engineering}}
  \bibinfo{volume}{26}, \bibinfo{number}{5} (\bibinfo{year}{2021}),
  \bibinfo{pages}{1--38}.
\newblock


\bibitem[\protect\citeauthoryear{Runeson}{Runeson}{2003}]%
        {runeson2003using}
\bibfield{author}{\bibinfo{person}{Per Runeson}.}
  \bibinfo{year}{2003}\natexlab{}.
\newblock \showarticletitle{Using students as experiment subjects--an analysis
  on graduate and freshmen student data}. In
  \bibinfo{booktitle}{\emph{Proceedings of the International Conference on
  Empirical Assessment in Software Engineering}}. \bibinfo{pages}{95--102}.
\newblock


\bibitem[\protect\citeauthoryear{Saini, Duan, and Paruchuri}{Saini
  et~al\mbox{.}}{2008}]%
        {saini2008threat}
\bibfield{author}{\bibinfo{person}{Vineet Saini}, \bibinfo{person}{Qiang Duan},
  {and} \bibinfo{person}{Vamsi Paruchuri}.} \bibinfo{year}{2008}\natexlab{}.
\newblock \showarticletitle{Threat modeling using attack trees}.
\newblock \bibinfo{journal}{\emph{Journal of Computing Sciences in Colleges}}
  \bibinfo{volume}{23}, \bibinfo{number}{4} (\bibinfo{year}{2008}),
  \bibinfo{pages}{124--131}.
\newblock


\bibitem[\protect\citeauthoryear{Salman, Misirli, and Juristo}{Salman
  et~al\mbox{.}}{2015}]%
        {salman2015students}
\bibfield{author}{\bibinfo{person}{Iflaah Salman}, \bibinfo{person}{Ayse~Tosun
  Misirli}, {and} \bibinfo{person}{Natalia Juristo}.}
  \bibinfo{year}{2015}\natexlab{}.
\newblock \showarticletitle{Are students representatives of professionals in
  software engineering experiments?}. In \bibinfo{booktitle}{\emph{Proceedings
  of the International Conference on Software Engineering-Volume 1}}. IEEE
  Press, \bibinfo{pages}{666--676}.
\newblock


\bibitem[\protect\citeauthoryear{Shostack}{Shostack}{2014}]%
        {shostack2014threat}
\bibfield{author}{\bibinfo{person}{Adam Shostack}.}
  \bibinfo{year}{2014}\natexlab{}.
\newblock \bibinfo{booktitle}{\emph{{Threat Modeling: Designing for
  Security}}}.
\newblock \bibinfo{publisher}{John Wiley \& Sons}. 590 pages.
\newblock
\showISBNx{978-1-118-80999-0}


\bibitem[\protect\citeauthoryear{Spichkova, Schmidt, and Trubiani}{Spichkova
  et~al\mbox{.}}{2017}]%
        {spichkova2017role}
\bibfield{author}{\bibinfo{person}{Maria Spichkova}, \bibinfo{person}{Heinz
  Schmidt}, {and} \bibinfo{person}{Catia Trubiani}.}
  \bibinfo{year}{2017}\natexlab{}.
\newblock \showarticletitle{Role of women in software architecture: an attempt
  at a systematic literature review}. In \bibinfo{booktitle}{\emph{Proceedings
  of the 11th European Conference on Software Architecture: Companion
  Proceedings}}. \bibinfo{pages}{31--34}.
\newblock


\bibitem[\protect\citeauthoryear{Thomas, Joseph, Williams, Burge,
  et~al\mbox{.}}{Thomas et~al\mbox{.}}{2018}]%
        {thomas2018speaking}
\bibfield{author}{\bibinfo{person}{Jakita~O Thomas}, \bibinfo{person}{Nicole
  Joseph}, \bibinfo{person}{Arian Williams}, \bibinfo{person}{Jamika Burge},
  {et~al\mbox{.}}} \bibinfo{year}{2018}\natexlab{}.
\newblock \showarticletitle{Speaking truth to power: Exploring the
  intersectional experiences of Black women in computing}. In
  \bibinfo{booktitle}{\emph{2018 Research on Equity and Sustained Participation
  in Engineering, Computing, and Technology (RESPECT)}}. IEEE,
  \bibinfo{pages}{1--8}.
\newblock


\bibitem[\protect\citeauthoryear{Tuma, Calikli, and Scandariato}{Tuma
  et~al\mbox{.}}{2018}]%
        {tuma2018threat}
\bibfield{author}{\bibinfo{person}{Katja Tuma}, \bibinfo{person}{G{\"{u}}l
  Calikli}, {and} \bibinfo{person}{Riccardo Scandariato}.}
  \bibinfo{year}{2018}\natexlab{}.
\newblock \showarticletitle{Threat analysis of software systems: A systematic
  literature review}.
\newblock \bibinfo{journal}{\emph{Journal of Systems and Software}}
  \bibinfo{volume}{144} (\bibinfo{year}{2018}), \bibinfo{pages}{275--294}.
\newblock


\bibitem[\protect\citeauthoryear{Tuma, Sandberg, Thorsson, Widman, Herpel, and
  Scandariato}{Tuma et~al\mbox{.}}{2021}]%
        {tuma2021finding}
\bibfield{author}{\bibinfo{person}{Katja Tuma}, \bibinfo{person}{Christian
  Sandberg}, \bibinfo{person}{Urban Thorsson}, \bibinfo{person}{Mathias
  Widman}, \bibinfo{person}{Thomas Herpel}, {and} \bibinfo{person}{Riccardo
  Scandariato}.} \bibinfo{year}{2021}\natexlab{}.
\newblock \showarticletitle{Finding security threats that matter: Two
  industrial case studies}.
\newblock \bibinfo{journal}{\emph{Journal of Systems and Software}}
  \bibinfo{volume}{179} (\bibinfo{year}{2021}), \bibinfo{pages}{111003}.
\newblock


\bibitem[\protect\citeauthoryear{Tuma and Scandariato}{Tuma and
  Scandariato}{2018}]%
        {tuma2018two}
\bibfield{author}{\bibinfo{person}{Katja Tuma} {and} \bibinfo{person}{Riccardo
  Scandariato}.} \bibinfo{year}{2018}\natexlab{}.
\newblock \showarticletitle{Two Architectural Threat Analysis Techniques
  Compared}. In \bibinfo{booktitle}{\emph{Proceedings of the European
  Conference on Software Architecture (ECSA)}}. Springer,
  \bibinfo{pages}{347--363}.
\newblock


\bibitem[\protect\citeauthoryear{Uhlmann and Cohen}{Uhlmann and Cohen}{2007}]%
        {uhlmann2007think}
\bibfield{author}{\bibinfo{person}{Eric~Luis Uhlmann} {and}
  \bibinfo{person}{Geoffrey~L Cohen}.} \bibinfo{year}{2007}\natexlab{}.
\newblock \showarticletitle{“I think it, therefore it’s true”: Effects of
  self-perceived objectivity on hiring discrimination}.
\newblock \bibinfo{journal}{\emph{Organizational Behavior and Human Decision
  Processes}} \bibinfo{volume}{104}, \bibinfo{number}{2}
  (\bibinfo{year}{2007}), \bibinfo{pages}{207--223}.
\newblock


\bibitem[\protect\citeauthoryear{Van~der Lee and Ellemers}{Van~der Lee and
  Ellemers}{2015}]%
        {van2015gender}
\bibfield{author}{\bibinfo{person}{Romy Van~der Lee} {and}
  \bibinfo{person}{Naomi Ellemers}.} \bibinfo{year}{2015}\natexlab{}.
\newblock \showarticletitle{Gender contributes to personal research funding
  success in The Netherlands}.
\newblock \bibinfo{journal}{\emph{Proceedings of the National Academy of
  Sciences}} \bibinfo{volume}{112}, \bibinfo{number}{40}
  (\bibinfo{year}{2015}), \bibinfo{pages}{12349--12353}.
\newblock


\bibitem[\protect\citeauthoryear{van~der Lee and Ellemers}{van~der Lee and
  Ellemers}{2018}]%
        {van2018perceptions}
\bibfield{author}{\bibinfo{person}{Romy van~der Lee} {and}
  \bibinfo{person}{Naomi Ellemers}.} \bibinfo{year}{2018}\natexlab{}.
\newblock \showarticletitle{Perceptions of gender inequality in academia:
  Reluctance to let go of individual merit ideology}.
\newblock In \bibinfo{booktitle}{\emph{Belief Systems and the Perception of
  Reality}}. \bibinfo{publisher}{Routledge}, \bibinfo{pages}{63--78}.
\newblock


\bibitem[\protect\citeauthoryear{Van~Dijk, Van~Engen, and
  Van~Knippenberg}{Van~Dijk et~al\mbox{.}}{2012}]%
        {van2012defying}
\bibfield{author}{\bibinfo{person}{Hans Van~Dijk}, \bibinfo{person}{Marloes~L
  Van~Engen}, {and} \bibinfo{person}{Daan Van~Knippenberg}.}
  \bibinfo{year}{2012}\natexlab{}.
\newblock \showarticletitle{Defying conventional wisdom: A meta-analytical
  examination of the differences between demographic and job-related diversity
  relationships with performance}.
\newblock \bibinfo{journal}{\emph{Organizational Behavior and Human Decision
  Processes}} \bibinfo{volume}{119}, \bibinfo{number}{1}
  (\bibinfo{year}{2012}), \bibinfo{pages}{38--53}.
\newblock


\bibitem[\protect\citeauthoryear{Van~Landuyt and Joosen}{Van~Landuyt and
  Joosen}{2021}]%
        {van2021descriptive}
\bibfield{author}{\bibinfo{person}{Dimitri Van~Landuyt} {and}
  \bibinfo{person}{Wouter Joosen}.} \bibinfo{year}{2021}\natexlab{}.
\newblock \showarticletitle{A descriptive study of assumptions in STRIDE
  security threat modeling}.
\newblock \bibinfo{journal}{\emph{Software and Systems Modeling}}
  (\bibinfo{year}{2021}), \bibinfo{pages}{1--18}.
\newblock


\bibitem[\protect\citeauthoryear{Vorvoreanu, Zhang, Huang, Hilderbrand,
  Steine-Hanson, and Burnett}{Vorvoreanu et~al\mbox{.}}{2019}]%
        {vorvoreanu2019gender}
\bibfield{author}{\bibinfo{person}{Mihaela Vorvoreanu}, \bibinfo{person}{Lingyi
  Zhang}, \bibinfo{person}{Yun-Han Huang}, \bibinfo{person}{Claudia
  Hilderbrand}, \bibinfo{person}{Zoe Steine-Hanson}, {and}
  \bibinfo{person}{Margaret Burnett}.} \bibinfo{year}{2019}\natexlab{}.
\newblock \showarticletitle{From gender biases to gender-inclusive design: An
  empirical investigation}. In \bibinfo{booktitle}{\emph{Proceedings of the
  2019 CHI Conference on Human Factors in Computing Systems}}.
  \bibinfo{pages}{1--14}.
\newblock


\bibitem[\protect\citeauthoryear{Wang and Wagner}{Wang and Wagner}{2018}]%
        {wang2018groupthink}
\bibfield{author}{\bibinfo{person}{Yang Wang} {and} \bibinfo{person}{Stefan
  Wagner}.} \bibinfo{year}{2018}\natexlab{}.
\newblock \showarticletitle{On groupthink in safety analysis: An industrial
  case study}. In \bibinfo{booktitle}{\emph{Proceedings of the 40th
  International Conference on Software Engineering: Software Engineering in
  Practice}}. \bibinfo{pages}{266--275}.
\newblock


\bibitem[\protect\citeauthoryear{Waring}{Waring}{2015}]%
        {waring2015managerial}
\bibfield{author}{\bibinfo{person}{Alan Waring}.}
  \bibinfo{year}{2015}\natexlab{}.
\newblock \showarticletitle{Managerial and non-technical factors in the
  development of human-created disasters: A review and research agenda}.
\newblock \bibinfo{journal}{\emph{Safety science}}  \bibinfo{volume}{79}
  (\bibinfo{year}{2015}), \bibinfo{pages}{254--267}.
\newblock


\bibitem[\protect\citeauthoryear{Wright, Bolger, and Rowe}{Wright
  et~al\mbox{.}}{2002}]%
        {wright2002empirical}
\bibfield{author}{\bibinfo{person}{George Wright}, \bibinfo{person}{Fergus
  Bolger}, {and} \bibinfo{person}{Gene Rowe}.} \bibinfo{year}{2002}\natexlab{}.
\newblock \showarticletitle{An empirical test of the relative validity of
  expert and lay judgments of risk}.
\newblock \bibinfo{journal}{\emph{Risk Analysis: An International Journal}}
  \bibinfo{volume}{22}, \bibinfo{number}{6} (\bibinfo{year}{2002}),
  \bibinfo{pages}{1107--1122}.
\newblock


\bibitem[\protect\citeauthoryear{Wynn and Correll}{Wynn and Correll}{2017}]%
        {wynn2017gendered}
\bibfield{author}{\bibinfo{person}{Alison~T Wynn} {and}
  \bibinfo{person}{Shelley~J Correll}.} \bibinfo{year}{2017}\natexlab{}.
\newblock \showarticletitle{Gendered perceptions of cultural and skill
  alignment in technology companies}.
\newblock \bibinfo{journal}{\emph{Social Sciences}} \bibinfo{volume}{6},
  \bibinfo{number}{2} (\bibinfo{year}{2017}), \bibinfo{pages}{45}.
\newblock


\end{thebibliography}
\end{document}